\documentclass{iau}

\usepackage{amsmath}
\usepackage{graphicx}
\usepackage{multirow}
\usepackage{subcaption}

\begin{document}

\lefttitle{Hendrik Mueller}
\righttitle{To CLEAN or not to CLEAN: Data Processing in the ngVLA era}

\jnlPage{1}{7}
\jnlDoiYr{2021}
\doival{10.1017/xxxxx}

\aopheadtitle{Proceedings IAU Symposium}
\editors{eds}

\title{To CLEAN or not to CLEAN: Data Processing in the ngVLA era}

\author{Hendrik Mueller}
\affiliation{National Radio Astronomy Observatory, 1011 Lopezville Road, 87801 Socorro, New Mexico, USA}

\begin{abstract}
Radio interferometric imaging has long relied on the CLEAN algorithm, valued for its speed, robustness, and integration with calibration pipelines. However, next-generation facilities such as the ngVLA, SKA, and ALMA’s Wideband Sensitivity Upgrade will produce data volumes and dynamic ranges that exceed the scalability of traditional methods. CLEAN remains dominant due to its simplicity and accumulated expertise, yet its assumption of modeling the sky as point sources limits its ability to recover extended emission and hampers automation. We review CLEAN’s limitations and survey alternatives, including multiscale extensions, compressive sensing, Regularized Maximum Likelihood, Bayesian inference, and AI-driven approaches. Forward-modeling methods enable higher fidelity, flexible priors, and uncertainty quantification, albeit at greater computational cost. Hybrid approaches such as Autocorr-CLEAN, CG-CLEAN, and PolyCLEAN retain CLEAN’s workflow while incorporating modern optimization. We argue hybrids are best suited for the near term, while Bayesian and AI-based frameworks represent the long-term future of interferometric imaging.
\end{abstract}


\maketitle

\section{Introduction}

\begin{figure}[h]
\centering
\includegraphics[width=\linewidth]{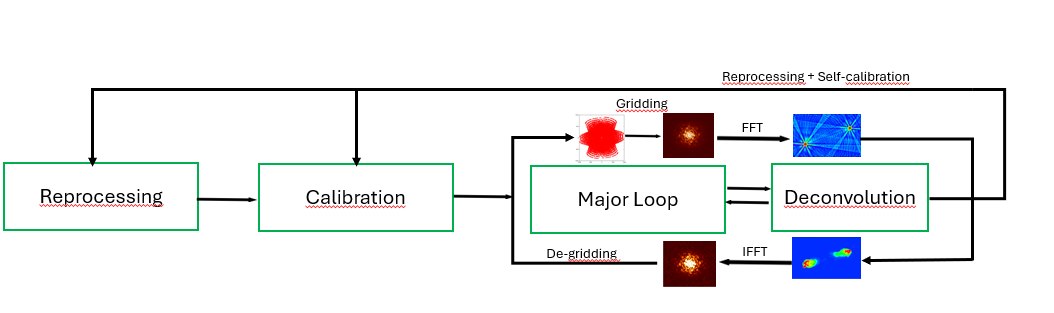}
\caption{Traditional data processing workflow.}
\label{fig:flow}
\end{figure}

Radio interferometry is a cornerstone of modern astrophysics, enabling studies of phenomena from supernova remnants to active galactic nuclei. Signals from multiple antennas are correlated, and by the van Cittert–Zernike theorem these correlations approximate Fourier components of the sky brightness \citep{Thompson2017}. In practice, instrumental effects (e.g., beam shapes, bandpasses, leakage, noise, digitization) and astrophysical corruptions (e.g., RFI, opacity, transients) complicate the measurement. Sparse Fourier sampling further turns image reconstruction into a deconvolution problem with strong sidelobe contamination.

Conventional processing couples calibration and imaging in iterative loops (Fig. \ref{fig:flow}). CLEAN and its multiscale and multifrequency variants \citep{Hogbom1974, Bhatnagar2004, Cornwell2008, Rau2011, Offringa2014, Offringa2017, Mueller2023a, Mueller2025a} remain the standard for minor-loop deconvolution, while major loops compare models to visibilities through gridding/degridding. High-fidelity imaging requires repeated refinements, making calibration, imaging, and quality control inseparable.  For the scope of this summary however, we limit the focus of the discussion mainly to the deconvolution stage. We refer the interested reader to \citet{Bhatnagar2025} for the description of the full algorithm architecture for radio interferometry data processing, and detailed considerations regarding its high performance computing implementation.

Beyond CLEAN, alternatives such as compressive sensing, Regularized Maximum Likelihood, Bayesian inference, and machine learning are rapidly advancing, and their systematic use in large-scale experiments marks a key frontier.

Next-generation facilities, including the ngVLA, SKA, and ALMA’s Wideband Sensitivity Upgrade (WSU), will produce unprecedented data volumes with higher dynamic range, broader bandwidths, and finer resolution. These advances place heavy demands on imaging pipelines for speed, accuracy, and automation. For the WSU, computing requirements reach $\sim 100$ Tflop/s, while the ngVLA will require $\sim 50$ Petaflop/s, with gridding as the dominant cost \citep{Bhatnagar2021, Kepley2024}. Even modest algorithmic improvements at this stage can yield significant savings. In the post-Moore era, scalability depends on massive parallelization, especially via GPUs. Quantum computing is not expected to solve the data size problem in radio interferometry in the near future \citep{Brunet2023}.

Future-ready pipelines require scalable algorithms that accelerate major loops through optimized GPU implementations and/or reduce their number through faster convergence. While higher sensitivity may necessitate advanced methods for super-resolution and extreme dynamic range, the majority of science cases should remain well served by CLEAN. The challenge is therefore to balance novel, high-fidelity methods with robust, efficient algorithms that scale to upcoming arrays, addressing both immediate upgrades and longer-term developments.

\section{Beyond Traditional CLEAN}

The CLEAN algorithm \citep{Hogbom1974} remains the workhorse of interferometric deconvolution. Its peak-finding and beam-subtraction procedure is fast, robust, and tightly integrated with calibration workflows. With heuristics such as automasking \citep{Kepley2020} and adaptive stopping \citep{Homan2024}, CLEAN is pragmatic, extensible, and, especially in multiscale variants, still competitive to newer approaches including compressive sensing, Bayesian inference, and AI \citep{Offringa2017, Bester2025}.

Nonetheless, increasing sensitivity and data complexity expose its limits. CLEAN’s assumption that the sky is a collection of point sources fails for extended or diffuse emission, while the method’s reliance on user input (masks, gains, stopping rules) restricts automation. It also enforces a rigid separation between model and image and offers no uncertainty quantification. Alternatives based on sparse modeling, Bayesian inference, and machine learning provide more flexible priors, potentially higher fidelity reconstructions, and principled error estimates.

Imaging algorithms are often divided into inverse modeling and forward modeling approaches \citep{eht2019d}. Inverse modeling refines CLEAN through improved basis functions and heuristics. MS-CLEAN and multifrequency variants \citep{Cornwell2008, Rau2011, Offringa2014, Offringa2017} use Gaussian components of varying scale sizes, Asp-CLEAN \citep{Bhatnagar2004,Hsieh2021,Hsieh2022a,Hsieh2022b} optimizes scale sizes adaptively, and wavelet-based methods such as DoB-CLEAN \citep{Mueller2023a} introduce elliptical difference-of-Gaussian bases. These approaches significantly improve fidelity while still remaining competitive with more advanced forward modeling paradigms \citep{Offringa2017,Hsieh2021,Hsieh2022b,Bester2025}.

Forward modeling, in contrast, formulates imaging as a direct optimization problem. Maximum entropy methods evolved into Regularized Maximum Likelihood, balancing data fidelity and multiple regularization terms \citep{Cornwell1985, Briggs1995}, and have been extensively developed in the Event Horizon Telescope context \citep{Chael2016, Akiyama2017, Chael2018}. More recent directions include multiobjective evolution to explore multimodality in degenerate solutions \citep{Mueller2023c, Mus2024b, Mus2024c}, compressive sensing with sparsity-promoting priors \citep{Starck2002, Starck2005, Wiaux2009, Carrillo2012, Carrillo2014, Dabbech2018, Mueller2022}, and Bayesian inference frameworks such as \texttt{resolve} \citep{Junklewitz2016, Knollmueller2019, Frank2021, Arras2021, Roth2024} or sampling-based methods in the EHT \citep{Broderick2020, Tiede2022}. AI-based techniques are also emerging rapidly \citep{Sun2022, Connor2022, Terris2023, Aghabiglou2024, Dabbech2024}.

Forward-modeling methods often achieve superior fidelity for complex structures, potentially offer an uncertainty quantification and can directly address nonlinear corruptions, e.g. closure-only imaging and polarization \citep{Chael2018, Mueller2023c, Mueller2024b}, joint calibration and imaging \citep{Arras2019, Kim2024}, direction-dependent effects \citep{Dabbech2021, Arras2021b, Roth2023}, or time-dynamic sources \citep{Bouman2017, Arras2022, Mueller2023b, Mus2024a, Mus2024c}. Their main drawback typically is computational cost, which remains a barrier to routine use.

Despite these advances, CLEAN persists as the de facto standard, valued for its speed, competitiveness of its modern multiscalar variants, and simply sufficiency for many science cases. Novel paradigms have shown improved performance in targeted tests, but large-scale validation across observatory pipelines is lacking, and deployment will require substantial additional development and commissioning.

Hybrid approaches offer strong near-term potential for pipelines supporting upcoming facilities such as SKA and ALMA’s WSU. They preserve the speed, robustness, and validation of CLEAN while replacing its suboptimal regularization, providing an evolutionary improvement without requiring new infrastructure. Notable examples include Autocorr-CLEAN \citep{Mueller2025a}, CG-CLEAN \citep{Mueller2025b}, and PolyCLEAN \citep{Jarret2025}. Even modern AI and Bayesian methods, such as R2D2 \citep{Aghabiglou2024} and fast-resolve \citep{Roth2024}, adopt major/minor loop structures for efficiency.

\begin{figure}
\centering
\includegraphics[width=\linewidth]{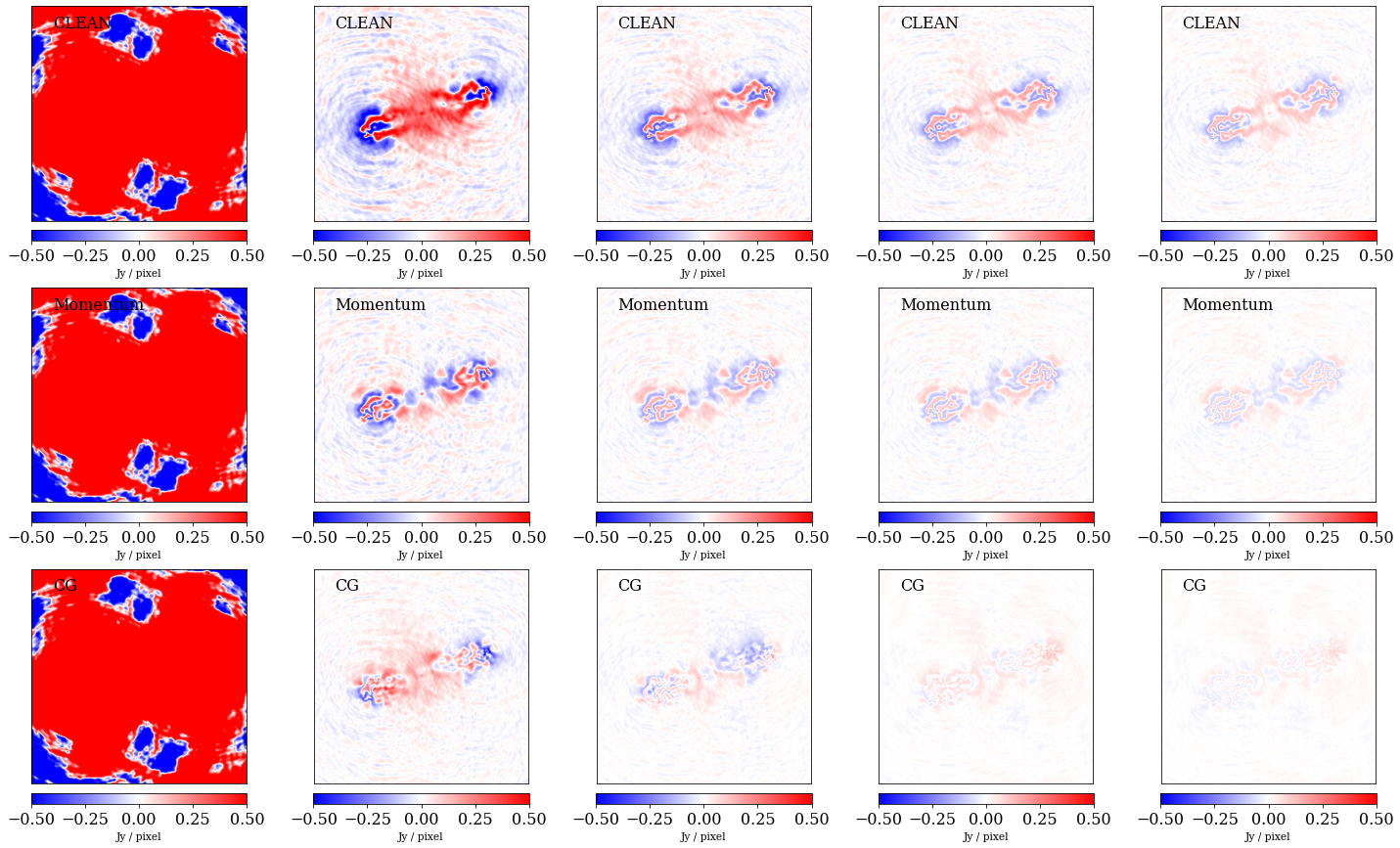}
\caption{Residual as a function of major loop iteration for CLEAN and two variants that couple CLEAN with optimization techniques (CG-CLEAN and Momentum-CLEAN) for Cygnus A. Image adapted from \citet{Mueller2025b}.}
\label{fig:residuals_cg}
\end{figure}

PolyCLEAN integrates CLEAN with an explicit LASSO formulation, while CG-CLEAN interprets the major loop as a gradient step and the minor loop as a Hessian inversion, casting the process into a conjugate gradient scheme. This accelerates convergence and drives residuals toward a noise-like distribution more efficiently than traditional CLEAN \citep{Mueller2025b}. We show the residuals as a function of number of major loops exemplary in Fig. \ref{fig:residuals_cg}.

\begin{figure}
\centering
\includegraphics[width=\linewidth]{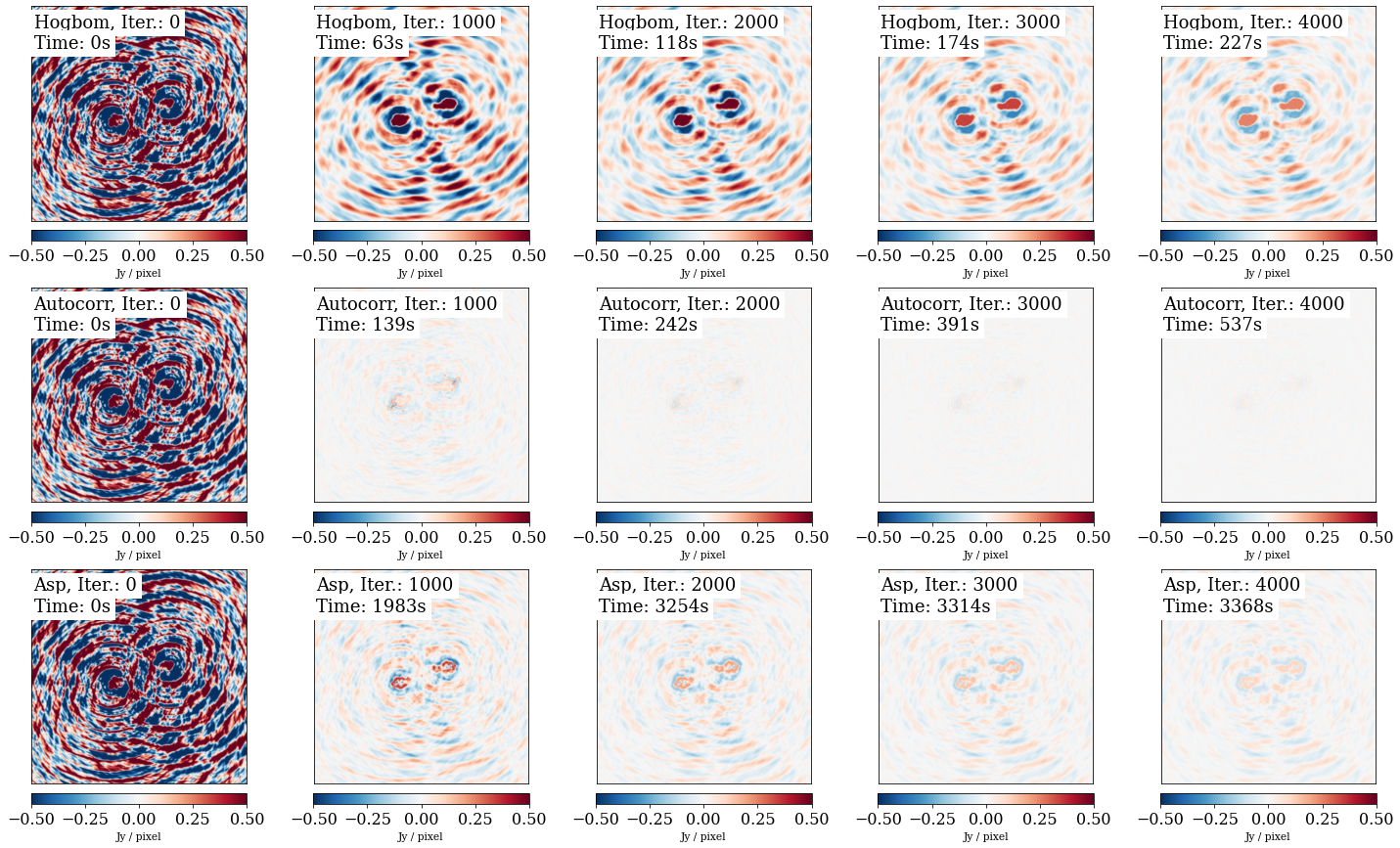}
\caption{Residual as a function of minor loop iteration and time for CLEAN (upper row), Autocorr-CLEAN (middle row) and Asp-CLEAN (bottom row) for Cygnus A. Image adapted from \citet{Mueller2025a}.}
\label{fig:residuals_auto}
\end{figure}

Other hybrids focus on enhancing multiscale representations. Asp-CLEAN \citep{Bhatnagar2004} and DoB-CLEAN \citep{Mueller2023a} improve accuracy by introducing flexibility in scale size or ellipticity, extending the gains of MS-CLEAN \citep{Cornwell2008}. Autocorr-CLEAN instead clusters components via residual autocorrelation, reducing iteration counts and improving convergence by up to an order of magnitude while remaining CASA-compatible. We show the residual of Autocorr-CLEAN and Asp-CLEAN as a function of number of minor loop iterations in Fig. \ref{fig:residuals_auto}, demonstrating the significant improvements of latter two over traditional CLEAN. Further medium-term developments may include subgridding techniques, which selectively only grid visibilities based on component scale to achieve additional computational savings. 

\begin{figure}
\centering
\begin{subfigure}[b]{0.5\textwidth}
\includegraphics[width=\linewidth]{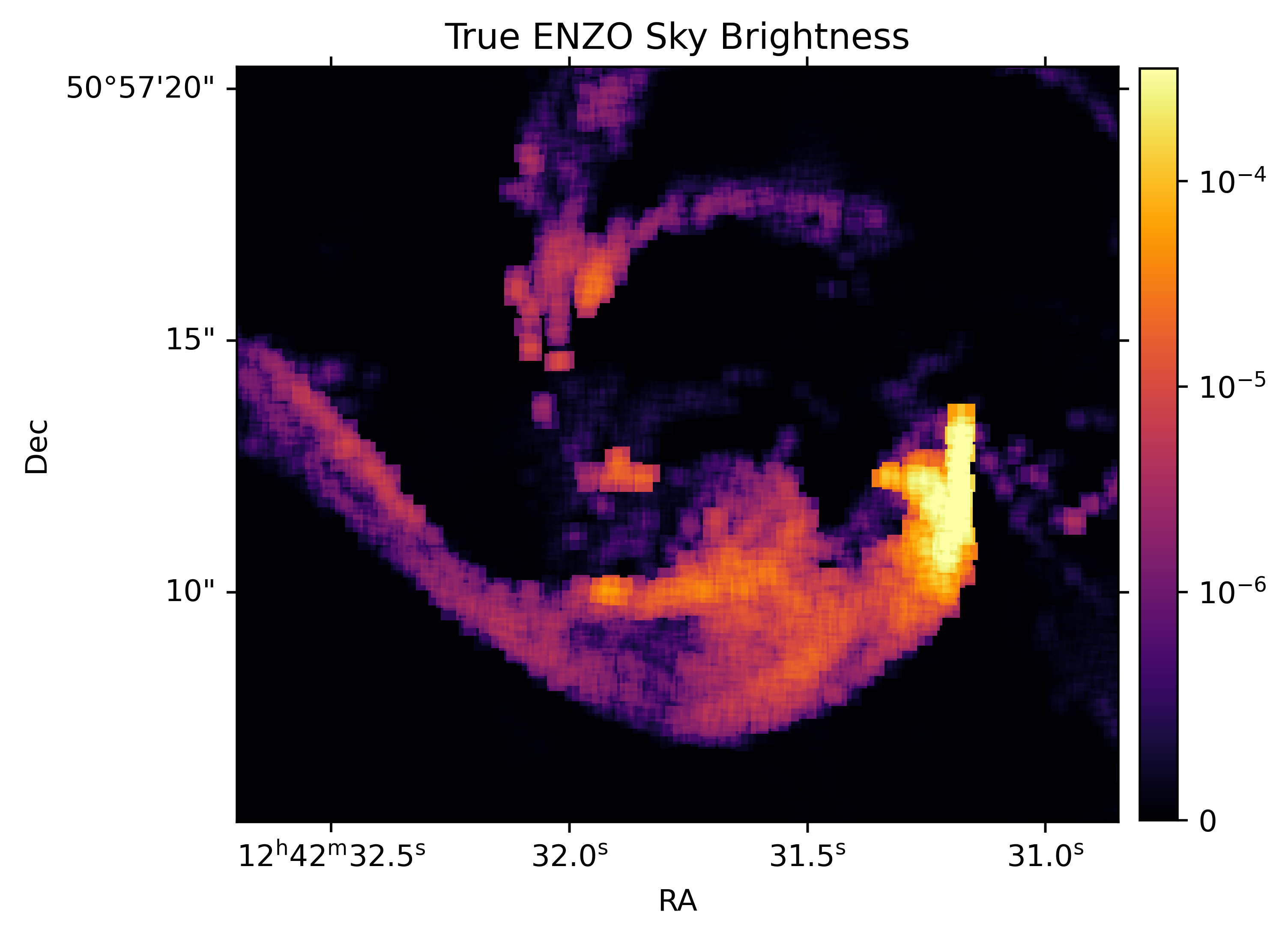}
\caption{Ground truth}
\end{subfigure}
\begin{subfigure}[b]{1\textwidth}
\includegraphics[width=\linewidth]{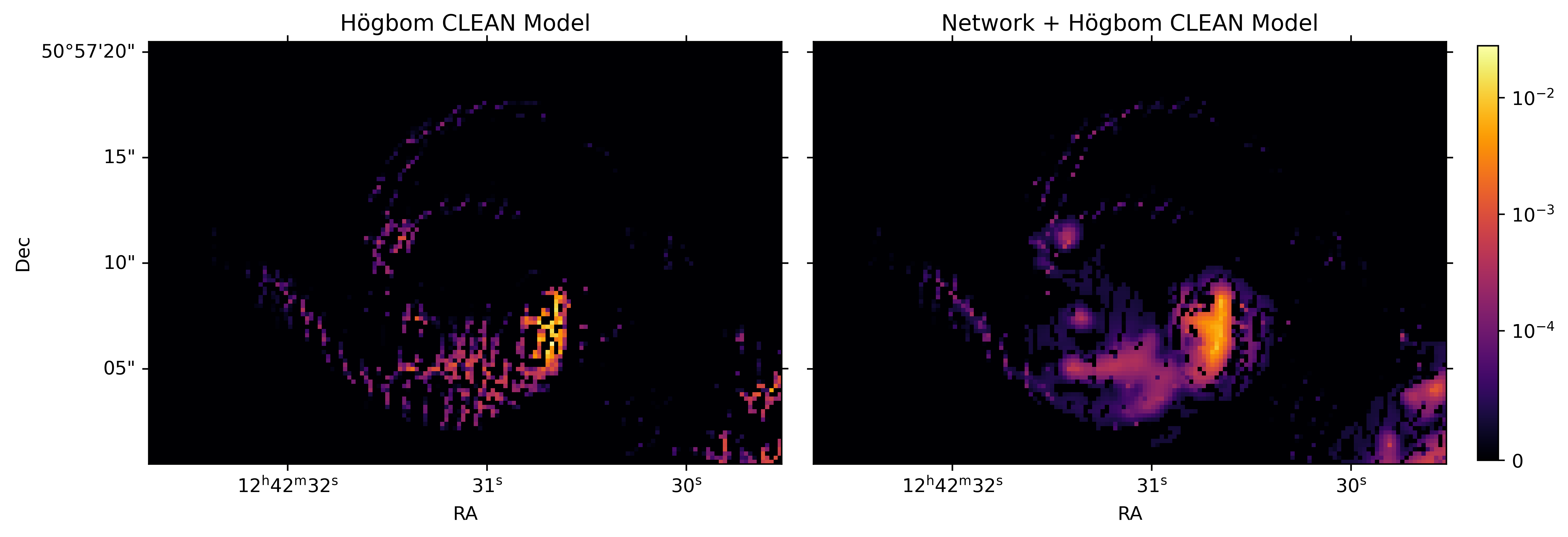}
\caption{Reconstruction}
\end{subfigure}
\caption{Reconstruction with traditional CLEAN and with a network transforming CLEAN components into a multiscale representation. Images are adapted from \citet{Fiedler2025}.}
\label{fig:AI}
\end{figure}

\section{Conclusions and Future AI Outlook}
The question of whether \textit{to CLEAN or not to CLEAN} is ultimately one of practicality. CLEAN is not obsolete, but its limitations become clearer as we push toward higher sensitivity, resolution, and data volume. Incremental improvements and hybrid extensions, coupled with modern computational techniques, provide a bridge until more advanced paradigms are ready for large-scale deployment.

In the near term, hybrid frameworks, such as Autocorr-CLEAN or convex optimization–based variants, offer faster convergence and higher dynamic range while retaining CLEAN’s reliability and simplicity. Looking further ahead, Bayesian and AI-driven methods are poised to redefine interferometric imaging, offering flexible priors, principled uncertainty quantification, and improved fidelity.

AI is expected to play a key role throughout this transition. Beyond plug-and-play reconstructions that exploit learned low-dimensional priors, AI can already enhance traditional workflows, for example, by learning policies to trigger calibration and imaging tasks \citep{Kirk2024} or by transforming Hogbom CLEAN components into multiscale representations \citep{Fiedler2025}. We demonstrate this approach in Fig. \ref{fig:AI}, where a network trained on single-scale and MS-CLEAN representations transforms single-scale images into multiscale ones. This overcomes the main limitation of the former and achieves the accuracy of the latter with minimal computational overhead and bias from the training set \citep{Fiedler2025}.

The future of radio interferometry will be shaped by our ability to balance accuracy, scalability, and practicality, ensuring imaging methods evolve in step with the demands of next-generation telescopes.

\bibliography{bibliography}{}
\bibliographystyle{aa}

\end{document}